\documentclass[german,english]{bvm2020}

\def\articlenumber{0000}

\date{}
\title{Analyzing an Imitation Learning Network for Fundus Image Registration Using a Divide-and-Conquer Approach}

\titlerunning{Fundus Image Registration}

\author{Siming~Bayer$^1$, Xia~Zhong$^1$, Weilin~Fu$^1$, Nishant~Ravikumar$^2$, Andreas~Maier$^1$}

\authorrunning{Bayer et al.}

\institute{%
$^1$Pattern Recognition Lab, FAU Erlangen-Nuremberg\\
$^2$CISTIB, School of Computing and School of Medicin, University of Leeds}

\email{siming.bayer@fau.de}

\begin{document}

%
\selectlanguage{english}

\maketitle

\begin{abstract}
Comparison of microvascular circulation on fundoscopic images is a non-invasive clinical indication for the diagnosis and monitoring of diseases, such as diabetes and hypertensions.
The differences between intra-patient images can be assessed quantitatively by registering serial acquisitions.
Due to the variability of the images (i.e. contrast, luminosity) and the anatomical changes of the retina, the registration of fundus images remains a challenging task. 
Recently, several deep learning approaches have been proposed to register fundus images in an end-to-end fashion, achieving remarkable results.
However, the results are difficult to interpret and analyze. 
In this work, we propose an imitation learning framework for the registration of 2D color funduscopic images for a wide range of applications such as disease monitoring, image stitching and super-resolution. 
We follow a divide-and-conquer approach to improve the interpretability of the proposed network, and analyze both the influence of the input image and the hyperparameters on the registration result.
The results show that the proposed registration network reduces the initial target registration error up to 95\%.

\end{abstract}

\section{Introduction}
Retina blood vessels and their morphological features are important biomarkers for non-invasive monitoring of chronic and age-related diseases, such as diabetic retinopathy, glaucoma, or macular edema. 
In order to analyse the progression of diseases, image registration techniques are used to conduct longitudinal studies by comparing serially acquired intra-patient funducsopic images.
Moreover, acquisitions from different viewpoints can be fused into one single image containing panoramic information of the retina.
However, registration of retinal images remains a challenging task due to various factors, such as uneven illumination of textureless regions, large variety of the viewpoints on serially acquired fundus images and pathological changes to the anatomy of the retina.

A detailed review~\cite{2815-01} of fundus image registration techniques shows that feature-based approaches are more frequently applied to resolve this task. 
In~\cite{2815-02}, a graph matching method is combined with ICP to find correspondences between vascular bifurcations and register retinal vessels. 
A comparison of feature-based retina image registration algorithms using bifurcations, SIFT or SURF as features is presented in~\cite{2815-03}. 
Those methods are accurate and are easy to interpret. 
A major limitation of conventional feature-based method is, that it consists of time consuming optimization methods. 
In order to address this limitation, \cite{2815-04} proposed a generative adverserial network (GAN) for the estimation the final dense deformation field. 
However, the network is only trained and tested on data, where the image pairs have large overlapping area.
Moreover, the entire registration pipeline is mapped into one single network, and operates as a black-box. 
This makes it difficult to interpret and to analyze. 

Recent advances towards better understanding of deep learning (DL) networks using precision learning~\cite{2815-05} encourage the decomposition of an end-to-end DL framework into known operators.
Previously, \cite{2815-06} proposed an imitation learning network for intra-operative brain shift compensation where the optimal displacement vectors of landmarks defined in the `source' image domain are predicted directly from the underlying image pair to be registered (in order to map them to the `target' image domain).
In this work, we use a modified version of this network and propose a generalized method for feature-based fundus image registration. 
The image pairs used for training and evaluation of the proposed network vary in their appearance, and are used for different applications.
In order to analyze the image registration network, we apply the concept of precision learning, following a divide-and-conquer strategy proposed in~\cite{2815-07} for the identification of the critical components which affects the accuracy of the limitation learning network.


\section{Materials and Methods}
\subsection{Fundus Image Registration Dataset (FIRE)}
\label{FIRE}
The FIRE dataset published in~\cite{2815-08} is a public database for retina image registration containing 134 pairs of intra-patient fundus images\footnote{Accessable via \url{https://www.ics.forth.gr/cvrl/fire/}}.
Moreover, ten homologous landmark pairs on the vessel bifurcation points are manually annotated for each image pairs.
All images are acquired in the same hospital with the same type of fundus camera. 
They have a resolution of $2912\times2912$ pixels and a field of view of $45^{\circ}\times45^{\circ}$.
In general, the image pairs can be divided into three categories with regard to the characteristics of the images:

    {\emph{Category \textsc{A}}} This category comprises 14 image pairs, which were collected during a longitudinal study with similar viewpoint. Due to pathological changes from disease progression image appearance between each pair of images varies greatly. 
    
    {\emph{Category \textsc{P}}} In this part, 49 intra-patient image pairs from different view points were captured for the purpose of image stitching. The anatomical differences within one image pair is small.
    
    {\emph{Category \textsc{S}}} In this group, 71 image pairs were acquired for the application of super-resolution. Therefore, the images within each image pair are captured from a similar viewpoint. 
    Anatomical differences or pathological changes within the retinal are invisible.

\subsection{Imitation Learning Network for Fundus Image Registration}
We employ imitation learning for the task of pair-wise registration of fundus images. 
The core idea of imitation learning is to train a network to mimic the behavior of the demonstrator based on current observations. 
In the case of pair-wise registration, the demonstrator predicts the optimal displacement vectors for known correspondences.

{\underline {\emph{Observation Encoding:}}}
We use both the spatial position of the landmarks and the image features associated with each landmark as our observation encoding.
The image features are defined as a concatenation of isotropic 2D patches around the landmarks on the input image. 
The 2D patches are extracted by resampling the original image with a isotropic spacial step size of $S$ and a patch size of $C \times C$.
For each point set, the normalized point spatial distribution is used as a part of observation encoding. 

{\underline{\emph{Demonstrator and Augmentation:}}}
As the point correspondence of the landmarks is known in the training data, the displacement vectors between the homologous points in a source-target image pair are directly considered as the demonstrator.
Each data set is augmented by varying the brightness and contrast relative to the original image.
Additionally, we also transform the original image and point-pairs using random affine transformations.

\begin{figure}[!ht]
  \centering
    \includegraphics[width=0.9\textwidth]{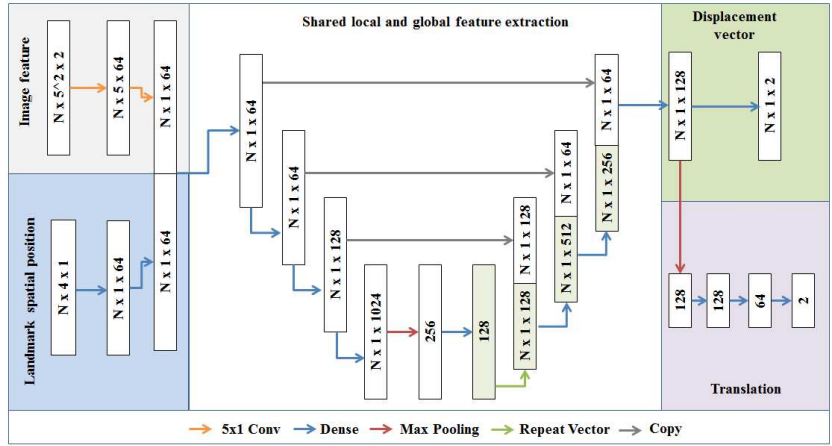}
    \caption{Architectures of the imitation network for the direct estimation of the displacement vectors on landmarks.}
    \label{fig:imitation_pipeline}
\end{figure}

{\underline{\emph{Network and loss:}}}
The network architecture of the proposed imitation network is illustrated in Fig.~\ref{fig:imitation_pipeline}.
The input to the network includes source and target images as well as their corresponding landmarks.
To facilitate the training of a robust registration network, a multi-task network is applied to predict the translation of the source image, and the displacement vectors of each source landmark, simultaneously.
The desired transformation of the source image can be estimated subsequently.
The same loss function as proposed in \cite{2815-06} is employed as a weighted loss for both tasks.

\subsection{Experimental Analysis of Imitation Learning Network}
Our network is trained and tested with the fundus images pairs introduced in Section~\ref{FIRE}.
A leave-one-out scheme was used to evaluate the registration performance of the imitation learning network on an independent held out test image-pair from the data set. While, within each fold of this leave-one-out scheme, the 133 image pairs were further split into training and validation sets using a ratio of $0.9:0.1$.
Both input images and the corresponding landmarks are augmented by creating 64 additional copies simultaneously.
The Adam optimizer with a learning rate of 0.001 was used throughout all experiments for training. 

In order to evaluate the generality and performance of the proposed method quantitatively, and understand the learning mechanism of the imitation learning architectures, e.g. to elucidate what the imitation network observes/learns, we conduct the following two experiments:

{\emph{Experiment I }}In this experiment, we analyze the influence of the appearance of the input images on the registration result.
Following a divide and conquer strategy proposed in \cite{2815-07}, we divide the fundus image registration pipeline into two parts, namely, image preprocessing (including vessel segmentation) and image registration.
Two different known operators are employed to resolve the task of image preprocessing prior to application of the imitation network.
The preprocessing and vessel enhancement techniques used in this study are validated and well understood in the literature. 
\begin{itemize}
    \item First, histogram equalization and a Laplacian of Gaussian filter (LoG) are applied. It is a straightforward analytical method with high noise sensitivity. 
    \item The second method for image preprocessing is adopted from~\cite{2815-07}, including a differentiable guided filter layer and an eight-scale Frangi-Net~\cite{2815-09}. 
    The CNN network pipeline for image preprocessing and segmentation is trained following the experiment setups in~\cite{2815-07}, and no further fine-tuning is conducted.
    Photographs in the FIRE database are standardized to (-1, 1) gray-scale images. To match the vessel diameters in the training data, images in FIRE are downsampled by a factor of four before fed into the network and upsampled after processing.
    Previous studies demonstrate, that the combination of a guided filter layer and Frangi-Net is a DL counterpart of the guided filter and the Frangi filter, on par with U-Net in terms of vessel segmentation accuracy. 
    However, it has fewer parameters and higher interpretability due to the use of known operators within the learning framework.
\end{itemize}
In both cases, the output of the image preprocessing block is a gray-scale image with enhanced vessel structures, which can be used as the input of the imitation network. Hereby, $C$ and $S$ are fixed to 20 and 40.


{\emph{Experiment II}} We use guided filter layer and Frangi-Net to preprocess the input image. 
The hyperparameters of the imitation network $C$ and $S$ are modified. 
$C$ is fixed to $20$ and $40$, whilst $S$ is changed. 
Here, the aim is to evaluate the influence of the hyperparameters on the registration result.

\section{Results}
\begin{figure}[!ht]
  \centering
    \includegraphics[width=\textwidth]{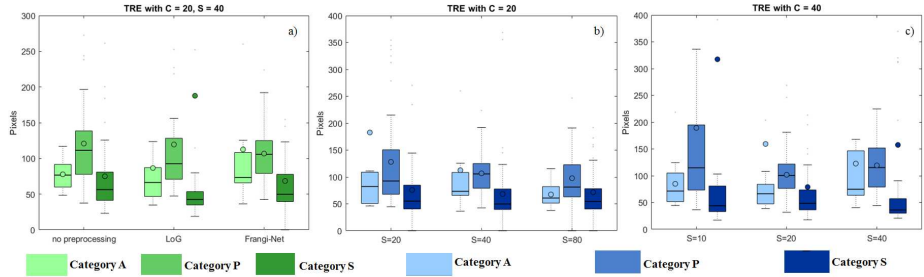}
    \caption{TRE between landmark positions predicted from the source points and the corresponding target points. The results of the \emph{Experiment I} are presented in Fig. a., whilst Fig. b. and Fig. c. show the results of \emph{Experiment II}.}
    \label{fig:res}
\end{figure}

Since annotated landmarks are provided as ground truth, we use target registration error (TRE) as the evaluation metric.
The result of both experiments are presented in Fig.\ref{fig:res}.
Initially, the TRE in pixels between the unregistered landmarks are $152.9\pm66.34$, $2518.9\pm948.22$, and $156.36\pm155.21$ for the image Category \textsc{A}, \textsc{P}, and \textsc{S}, respectively. 
 
\section{Discussion}
The quantitative results in Fig.~\ref{fig:res} shows our proposed network is able to recover 50\%, 95\%, and 75\% of the initial displacement for image Category A, P, and S respectively. 
The results of \emph{Experiment I} in Fig.~\ref{fig:res}a. indicates that input images preprocessed using Frangi-Net improve the registration performance of the network, for the image categories P and S, when $C=20$ and $S=40$. 
For the image Category A however, image preprocessing degrades the performance of the imitation network.
Considering the characteristics of the input image of each Category, we can draw the following conclusions: for input image pairs with similar appearance (i.e. categories P and S), the imitation network focuses and learns primarily from the vascular structures of the retina.
Thus, an accurate image preprocessing and segmentation method could benefit the overall performance of the registration step.
In Category A where the input images show a large variation in their appearance, the proposed network relies other image features besides the vasculature. 
Vessel enhancement and segmentation techniques such as LoG or Frangi-Net enhance vascular structures and suppress other information.
Therefore, the application of those techniques on the input images affect the performance of the proposed registration network for Category A negatively.

In the result of \emph{Experiment II} (Fig.~\ref{fig:res}b. and ~\ref{fig:res}c.) the impact of the hyperparameters $C$ and $S$ are demonstrated.
Small $C$ values were found to be more suitable to predict the displacement vectors of the landmarks in all three categories, when the input images are preprocessed and segmented using the guided filter layer and Frangi-Net. 
This effect is more visible in categories A and P, where the overlapping area of the input images are large, i.e. the initial distance of the landmarks are small.
In Category P, the results in both cases $C=20, S=40$ and $C=40, S=20$ are comparable.
Furthermore, they outperform other hyperparameter combinations.
To summarize, in \emph{Experiments II} $C$ can be considered as the resolution of the observation, whilst $S$ represents the extent of the entire observation.

A comprehensive comparison between the proposed network and iterative fundus image registration methods will be performed in future studies.
The registration accuracy of the proposed network could potentially be improved by automatically identifying the optimal values for the associated hyperparameters, namely, $C$ and $S$, based on the characteristics of the input image paris to be registration. This will also be investigated in future studies.


\bibliographystyle{bvm2020}

\bibliography{2815}
\end{document}